\documentclass{IEEEtran}

\usepackage{amsfonts}
\usepackage{amsmath}
\usepackage{amssymb}
\usepackage{mathptmx}
\usepackage{nicefrac}
\usepackage{url}
\usepackage{pifont}
\usepackage{fancyhdr}
\usepackage{multirow}
\usepackage{lineno}
\usepackage{cite}
\usepackage{graphicx}
\usepackage{tabularx}
\usepackage{color}

\definecolor{officialblue}{RGB}{0, 93, 170}					

\definecolor{purple}{RGB}{204, 0, 255}					

\definecolor{orange}{RGB}{247,127,0}						

\newcommand{\OMIT}[1]{} 

\begin{document}

\title{Lessons Learned from Development of a Software Tool to Support Academic Advising}

\author{\thanks{Manuscript received \today. This work is supported by the National Science Foundation, under grants CCF-1049360, ITR-0325063, and IIS-1107011.T. Dodson is supported by the Office of Naval Research Multidisciplinary University Research Initiative, Award Number 09-ONR-1115. Any opinions, findings, and conclusions or recommendations expressed in this material are those of the authors and do not necessarily reflect the views of the National Science Foundation or the Office of Naval Research.}\thanks{NICTA is funded by the Australian Government through the Department of Communications and the Australian Research Council through the ICT Centre of Excellence Program.}
		Nicholas Mattei\thanks{N. Mattei affiliated with NICTA, University of New South Wales, Neville Roach Laboratory, Level 4, 223 Anzac Pde., Kensington NSW 2052 Australia. (Nicholas.mattei@nicta.com.au); +61 2 8306 0464 .},
		Thomas Dodson\thanks{T. Dodson affiliated with Department of Physics and Astronomy, University of Pennsylvania, Philadelphia, PA 19103 USA (tcdodson@gmail.com).}, 
		Joshua T. Guerin\thanks{J. T. Guerin affiliated with Department of Computer Science, University of Tennessee at Martin, Martin, TN 38238 (jguerin@utm.edu).}, 
		Judy Goldsmith\thanks{J. Goldsmith affiliated with Department of Computer Science, University of Kentucky, Lexington, KY 40506 (goldsmit@cs.uky.edu).}, 
		Joan M. Mazur\thanks{J. M. Mazur affiliated with Department of Education, University of Kentucky, Lexington, KY 40506 (jmazur@uky.edu).}}

\maketitle

\begin{abstract}
We detail some lessons learned while designing and testing a decision-theoretic 
advising support
tool for undergraduates at a large state university.  Between 2009 and 2011 we conducted
two surveys of over 500 students in
  multiple majors and colleges.  These surveys asked students detailed questions
 about their preferences concerning course selection,
advising, and career paths.  We present data from this study
which may be helpful for faculty and staff who 
advise undergraduate students.  We find that advising support software tools 
can augment the student-advisor relationship, particularly in terms of course planning, 
but cannot and should not replace in-person advising.
\end{abstract}


\IEEEpeerreviewmaketitle

\section{Introduction}

At the University of Kentucky,
students in both the College of Engineering and the College of Arts and Sciences 
are required to meet with an advisor 
each semester before signing up for the following semester's courses.
Advising duties are
split between faculty members and full-time administrative staff whose primary or secondary duties include
professional and academic advising.
The advisors have access to the students' transcripts and are expected to know 
the course offerings for future semesters, 
requirements of the undergraduate degrees, prerequisite chains for the department's 
courses, possible career opportunities,
and the courses that will best prepare students to meet their 
post-graduation goals both in industry and academia.  
Advisors should also be able to guide the students in selecting courses that are
best suited to their abilities and goals.  
Finally, advisors 
should be able to refer students to support services, including academic 
support, special needs services, and counseling. 
What makes advising challenging is the need to personalize advice for
full-time and part-time students, transfer students, and students changing majors after satisfying
some of their previous major's requirements.

Ideally the student and advisor
keep in regular contact; the advisor plays a
supporting role in the student's continued development and formulates short and long term
goals for the student based on their individual needs and interests.
The reality is that most students see their advisor
once per semester for 15 to 30 minutes, sometimes see a different advisor each semester, and
sometimes see multiple advisors who are not necessarily in communication with one another.
Advisors may have their own 
agenda. Some may want to make certain that particular courses have high enough enrollment, while
some may assume that 
they know what students want. Some of the advisors in our study get extremely 
high evaluations from both students and faculty---usually those
who take the time to talk with students and help them understand how to set and achieve 
suitable goals.

We have developed components of an automated advising support system to 
augment the advisor-student relationship.
By allowing students to explore course offerings, possible future scenarios, and the
probabilistic outcomes of those future scenarios, we hope  our system will allow
students to enter their mandatory advisor meetings more conversant in their options. 
This 
preparation
would allow the human advisor to spend less time on more formulaic aspects of advising, such as explaining the course offerings
and requirements, and more time on career counseling, 
student support, and goal clarification. In addition, the system 
would provide a tool for advisors to explore and 
evaluate options with the students.

In recent years, there has been enormous growth and innovation in available 
online education tools and modalities.  
There is ongoing work, both commercial and academic, in academic advising tools.  However, we
argue that the continued development of online advising tools has not
kept pace with development of course delivery, educational theory
about online education, or education evaluation systems.   In the process of designing an
advising support system, we have done preliminary surveys about what students want from advisors,
and what advisors wish to offer.  The results of those surveys are guiding our own development
of advising support tools, and we hope they will be useful to others engaged in similar
development.

We believe that some of our findings are particular to Engineering, 
Computer Science, or other majors which have a strong career focus within the curriculum.
Indeed, we saw a very strong bias in the student responses toward advising information
related to the effect a given course would have on their future careers. 
We also conjecture that students in technical majors are more comfortable with the use of computers
and online tools than the student body at large; though this distinction may become less important
as technology continues to saturate our day-to-day lives.

Additionally, the results of our user testing may have implications for the way that human advisors
interact with their students by means of text-based communications (i.e., e-mail).  Our advising support
system provides the user with specific recommendations based on the results of decision-theoretic
planning algorithms applied to models of student goals and statistical models of
student performance.
However, the presentation of this information is made by providing plaintext explanations or arguments
that explain \emph{why} a particular course of action is the best.  Our discussion of effective
text-based explanations may be of particular relevance in situations where e-mail is used as a
secondary (in the case of a traditional advising setting) or even primary form of communication
between advisors and advisees (in the growing settings of distance education and e-learning).

Finally, we note that while the explanations presented in this paper are specific to the surveyed programs at the University of 
Kentucky, 
the explanation system itself was designed to be domain-independent 
\cite{DodsonNaturalLanguage2011,DodsonNatural2013}.
That is, the algorithms which generate
the explanation do not depend on the specifics of the degree program, and the system was designed to permit modification
of the underlying statistical model in order to support varying levels of student autonomy found at different institutions.

\section{Background and Motivation}\label{sec:background}
Academic advising, for many, is a full time job that requires
as much commitment, preparation, and care as teaching.
There is significant research into the theory of advising by incorporating 
principled pedagogical goals into the advising process and providing 
practical directives for practitioners \cite{gordon2011academic,lowenstein2005if}.
The availability of high-quality advising services has been
identified as an area of great importance in higher education.
Frequent, high-quality academic advising has been shown to have a positive effect
on GPA, satisfaction in the advising process, perceived value of education, and
attrition rates (both directly and indirectly) \cite{KingAcademicAdvisingRetention1993,MetznerPerceivedQuality1989,LoweGivers2001,TitleyNeglectedDimension1982}.  Additionally, recent research has shown that
minority and otherwise disadvantaged students can benefit the most from
quality advising services \cite{econ-gap}.

The links between the availability and quality of academic advising services and
lowered attrition rates may be of particular consequence for Engineering and Computer Science,
where attrition rates are often particularly high.
The University of Illinois at Urbana-Champaign reported a first-year attrition rate of Computer Science majors
of approximately 25\% \cite{TaltonScavengerHunt2006} over a 5 year period.
A study by Moller-Wong and Eide \cite{ISU97Retention} revealed similar trends across Engineering disciplines ---
30\% of students tracked in the study had left college entirely within 5 years, while 55\% had either left school or moved to 
a non-engineering major. A more recent study conducted at Rowan University \cite{HartmanLeavingEngineering} --- a college 
which has embraced current ``best practices'' with respect to student retention, and rentention of 
female students in particular --- reported a retention
rate of roughly 89\%, without the usual gender gap. 
A single-cohort study at North Carolina State University found a
3-year attrition rate of 17\% for students in the 2002 cohort \cite{NCSURetention2002}, while another 
study of the 2003 cohort at a major Australian research university found that 35\% of
students had left engineering by the end of the 6-year study period \cite{godfrey2010retention}, with a 28.5\% 3-year retention rate.

Because students frequently come into engineering programs with key skills deficits and, 
particularly in the case of Computer Science,
incorrect preconceived notions about the actual focus of the discipline, it has been postulated that
academic advising may be of particular importance for mitigating attrition rates in computer science
\cite{BeaubouefHighAttrition05}. Hartman et al. \cite{HartmanLeavingEngineering} found that
students with lower mathematics SAT scores and less experience with
high school math and science classes were significantly more likely to leave the program, 
while students who participated in discipline-specific engineering organizations where significantly more
likely to be retained. 
Indeed, both of these are factors that can be addressed through pre-enrollment academic advising, 
by offering support to either direct students
to more appropriate programs or to resources that can address key skills deficits, while encouraging them
to become active in on-campus engineering organizations.

Recent research has also demonstrated that computer-based advising tools can
be used to consolidate and simplify complicated advising information for the student and
advisor, and that such systems can have a measurable impact on student satisfaction in the
advising process \cite{FeghaliWebBased2011}.

The perceived importance of advising support software in the academic advising experience is also
reflected by the
actions of colleges and universities, many of which currently provide access to online advising
tools of varying complexity.  Several large public universities in the US
license software from College Source, specifically their u.achieve product.
This
course requirement checking tool is integrated with the universities course signup system
to provide customized degree audits for students.
Concepts from algorithms developed for recommender systems have
been integrated into more advanced systems that play a more active role in the interaction between student
and advisor.
A program at Austin Peay State 
University can help students select courses
based on their predicted grades, elicited interests, and graduation requirements
\cite{advising:netflix:10}.  Other researchers
are working on advising support systems which use collaborative filtering algorithms \cite{CF-advising}.

While such systems are a step in the right direction, we argue that their efficacy may be limited
by two important factors.  First, while these systems provide recommendations about the next course of
action they are missing a critical idea --- \emph{explaining} the rationale behind their
recommendations.  Second, recommender systems and collaborative filtering systems
typically do not consider uncertainty of the outcomes of advising actions or the potential
long-term effects of this uncertainty.

The notion of explaining why a particular course has been recommended is unlikely to be foreign to
experienced academic advisors (e.g., you should take course \emph{X} to better prepare yourself for
course \emph{Y}).  At the moment, only a few existing systems attempt to
do this 
\cite{DodsonNaturalLanguage2011,DodsonNatural2013,khan:mse_book}.
Explanation is an important part of recommendation \cite{RecSysHandbook_Explanation}.
Involving users in dialogue can improve the probability that recommendations are considered 
valid and adopted \cite{mahmood:mdp_travel}.

The notion of uncertainty of outcomes is likely to present difficult reasoning challenges,
regardless of quality of advisor. 
Even for good advisors, asking humans to make plans in domains in which the 
outcomes of actions are uncertain 
(e.g., course selection for students) invites many cognitive biases.
Humans are demonstrably poor at reasoning with uncertainty and are subject to,
for example, framing bias \cite{tandk:uncertainty}. Explanation generated by an
automated tool which is not subject to these cognitive biases can, sometimes, help them to 
reason about possible outcomes \cite{shani:mdp_bookstore,mahmood:mdp_travel}.

Our system generates arguments
that are designed to convince the user of the ``goodness'' of the recommended action
based on the internal mathematical model of advising\footnote{Technical details of our system
\cite{DodsonNaturalLanguage2011,DodsonNatural2013}
and model building procedures 
\cite{guerin:tr:model,guerin2012academic}
can be found in our other publications about the system.}.
This model is designed to be robust
even in the face of uncertain actions, as is evident by the multiple possibilities evaluated in its explanations.
Our system presents, as a paragraph, an argument that tries to convince the student 
to take a particular course in the next semester---a system which considers multiple courses
per semester is a focus of future research.
The underlying policy can be tailored to the student's preferences 
and abilities. Also, a case-based
algorithm generates an argument by analogy to the past performance of other students, enhancing transparency and persuasiveness \cite{nugent:cbe}.
It attempts to convince the user to
adopt the recommendation by demonstrating that other students have taken the same course sequence and succeeded. Explanations take the following form:
 
{\small
\begin{quote}
The recommended action is taking \textit{Introduction to Program Design and Problem Solving}, generated by 
examining possible future courses. It is the optimal course with regards to your 
current grades and the courses available to you.
Our model indicates that this action 
will best prepare you for taking \textit{Introduction to Software Engineering} and taking \textit{Discrete Mathematics} in the future. Additionally,
it will prepare you for taking \textit{Algorithm Design and Analysis}.
Our database indicates that with either a grade of A or B in \textit{Introductory Computer Programming} or a grade of A or B in
\textit{Calculus II}, you are more likely to receive a grade of A or B in \textit{Introduction to Program Design and Problem Solving},
the recommended course. 
\end{quote}
}

An additional algorithm was added to later iterations of our software
\cite{DodsonNatural2013},
specifically 
addressing student concerns raised in the Explanation System Survey. The algorithm is a probabilistic case-based calculation 
of the time from the current state to a user-defined goal (e.g., graduation or passing a particular ``capstone'' course) and 
presents a sentence in the following form:

{\small
\begin{quote}
Past students have taken \emph{Software Engineering}
and accomplished their goal of achieving \emph{a passing grade in Algorithm Design and Analysis}
in four or fewer \emph{semesters} from the current state.
\end{quote}
}

The goal of this research is to explore
the impact of explanation on the adoption
of recommended courses of action in uncertain domains.  In order to understand what makes
a good explanation in our initial target domain, academic advising, 
we interviewed many students and advisors about features
that make advice  compelling, and what goes into student decisions
regarding course selection.

\section{Survey Results}
Our data were collected from two anonymous surveys during the 2009--2010 and 2011--2012 academic years.
The 2009--2010 survey was focused on identifying students' needs and attitudes, specifically about advising.
 The 2011--2012 survey was conducted after the construction of our system in 
 order to gauge the effectiveness of our explanations in an advising domain.  Unless otherwise noted,
 all data comes from the Explanation System Survey.

\textbf{Advising Attitudes and Needs Survey (AANS):}
Over the course of the Fall 2009 and Spring 2010 semesters
we surveyed approximately 326 students enrolled in the University of Kentucky's
Introduction to Computer Programming
course (the first course in our major sequence).  Because an introductory computer
programming course is required of all engineering majors within our
college, we received responses primarily from Computer Science students, with a smaller
representation from other Engineering disciplines --- including Civil, Computer, and Electrical Engineering ---
as well as Mathematics, Education, and Physics, with the remaining responses primarily listing
their major as ``Undeclared'' or ``Other.''

This survey was conducted prior to the development of our advising explanation system.
The survey was exploratory:  we sought to discover
whether there was a need for more advanced computational advising tools, 
and in what capacity such tools would serve.
Along with demographic information for classification purposes, 
we collected data regarding the frequency with which
students sought university advising services, why they sought out an advisor, 
whether they used the online tools provided
by the university, and how valuable they perceived their advising experiences to be.

\textbf{Explanation System Survey (ESS):}
After the development of our advising support system we conducted a large user study
encompassing both target users of our system and domain experts (advisors).  
In our target user survey we surveyed 65 students enrolled in introductory computer science courses (``CS'' group).  These 
courses are open to all students, so a variety of majors are represented including
computer science, computer engineering, electrical engineering, physics, math, and 
mechanical engineering.  We also surveyed 130 students enrolled in introductory
psychology courses (``PSY'' group), which are also
open to all students.  The students surveyed included primarily majored in
psychology, but also included biology, social work, family sciences, and
undecided majors. This variety
allows us to make more general
statements about the types of advice that different students would prefer.

The EES was limited to paper surveys. As our system becomes more robust we hope 
to use it in controlled, real-world settings with both students and advisors in order
to study its effectiveness.
Surveys were handed out
with narratives based on two fictional, but plausible students.  Both students
are about half-way through completing a minor in their respective course of study:
one student is doing very well (about a 3.5 GPA) and one is struggling (2.3 GPA).
Survey respondents were asked to evaluate the advice our system generated for these 
students. From the demographic portions of the survey we know that 
most (more than 75\%) of the students
who took the survey in CS and PSY were within 2 semesters 
(plus or minus) of the fictional students and, in general, had GPA's close to the 
the fictional high achieving student.

In our domain experts survey we, conducted a survey of 10 advisors 
in order to gain perspective on how domain experts feel about our system 
and to validate our results against their advice.  
The advisors were computer science faculty
advisors, general College of Engineering advisors, and 
staff advisors from the College of Arts and Sciences
advisors.

When we authored our study instrument we had a variety of study
goals in mind.  In addition to demographic information, 
we wanted to know when and where users would 
interact with our system,  what they thought about the advice
generated by our system, subjective user and expert assessments of our system on various features,
and what factors users and experts would want to add to our system.
We included questions regarding their perceptions of 
the advising process and specific factors affecting their decisions.  
We do not provide a full analysis of the survey results
in this paper. Instead, we are focused on the attitudes of students about advising and general
attitudes about automated course advising tools.  Additional results can be found in our
other papers on this topic 
\cite{DodsonNaturalLanguage2011,DodsonNatural2013}.

\subsection{Student Attitudes}
Overall, the survey validated our method of advising support. High levels of 
agreement are shown between the students' decision-making and the framing of 
the arguments generated by the model-based and case-based explanation system:
47 of 62 ($75\%$) in the CS group and 104 of 130 ($80\%$) 
in the PSY group indicated that they considered how past students in their situation 
performed and/or how a course would prepare them for future courses to be 
important when making a decision. 
The latter method corresponds exactly to our model-based method of explanation in terms of short-term utility, 
while the former corresponds to our case-based method of explanation.
The suitability of argument by analogy in this domain was also validated:
38 of 62 ($61\%$) in the CS group and 65 of 130 ($50\%$) in the 
PSY group indicated that they considered the performance of 
past students in their situation.

Other survey results highlight the ability of our system to support
the advisor-student relationship.
The students seemed to be very goal focused: 42 of 62 ($68\%$) 
in the CS group and 100 of 130 ($76\%$) in the PSY group 
responded that course requirements were an important factor in 
deciding what courses to take. The model which our tool uses
incorporates course requirements implicitly, and the version
presented in 
\cite{DodsonNatural2013} 
explicitly addresses student concerns about time to graduation --- a common student request in the ESS.

However, more than $50\%$ of the students 
in both groups who responded to these questions (38 students in the 
PSY group and 25 in the CS group) had concerns about subjective factors of courses.  
These concerns included how many projects were assigned, what 
the professor was like, and whether taking two particular courses 
concurrently make for a particularly difficult semester. While a more
complex model of student preferences could take some of these subjective
factors into account, this result, more than any other,
underscores the utility of our tool as an \emph{advising support
system} (rather than an \emph{advising system}), reducing the amount of time spent
discussing the more formulaic aspects of course selection.

\subsubsection{Predicted Usage Patterns}
Most students responded that they would use the system at 
home before and/or while talking to an advisor.  31 of 44 ($70\%$) in the CS group and 95 of 121 ($78\%$) 
in the PSY group responded that they would use the tool at home, while 14 of 44 ($32\%$) in the CS group and 
64 of 121 ($53\%$) in the PSY group responded that they would use the tool while talking to an advisor.
Students were allowed to select multiple responces, and overall $84\%$ in the CS group and $87\%$ in the
PSY group responded that they would use the tool either at home \emph{or} while talking to an advisor---the
intended use pattern.

Engineering and Computer Science students
seemed particularly wary of our model, and indicated that they would be less willing to use the 
tool, if available, than students in other disciplines. When students were asked if they would make use of the advising 
feature if it was integrated with our university's course requirement 
checking feature --- 24 of 44 ($55\%$) for CS and 88 of 120 ($73\%$) 
for PSY, responded that they would often or always use the 
recommendation feature.  Many of the students who expressed a preference for not using
the tool were worried that it did not take into account \emph{all} their preferences ---
the more technical among them asked many questions about how the model was built and 
directly questioned its ability to capture their particular preferences.
This again highlights the utility of an advising \emph{support} tool --- the students were interested in
using the tool as a rote course requirement checker and for gaining a feel of what courses to take.
However, they are more comfortable when a human advisor is in place
to support them and make the recommendations more personal.

There was a very small group of students, 
7 of 44 ($15\%$) for CS and 27 of 165 ($16\%$) for PSY, that 
said they would use our system \emph{instead of} talking to an advisor.
This seems to correspond with our observations that some students
view advising as a chore due to difficulties of scheduling time to meet 
an advisor. In fact, the relatively low percentage who
would choose to use completely automated advising is encouraging.

\begin{figure*}[ht]
  \begin{minipage}[t]{0.48\linewidth}
    \centering
   \includegraphics[height=5cm]{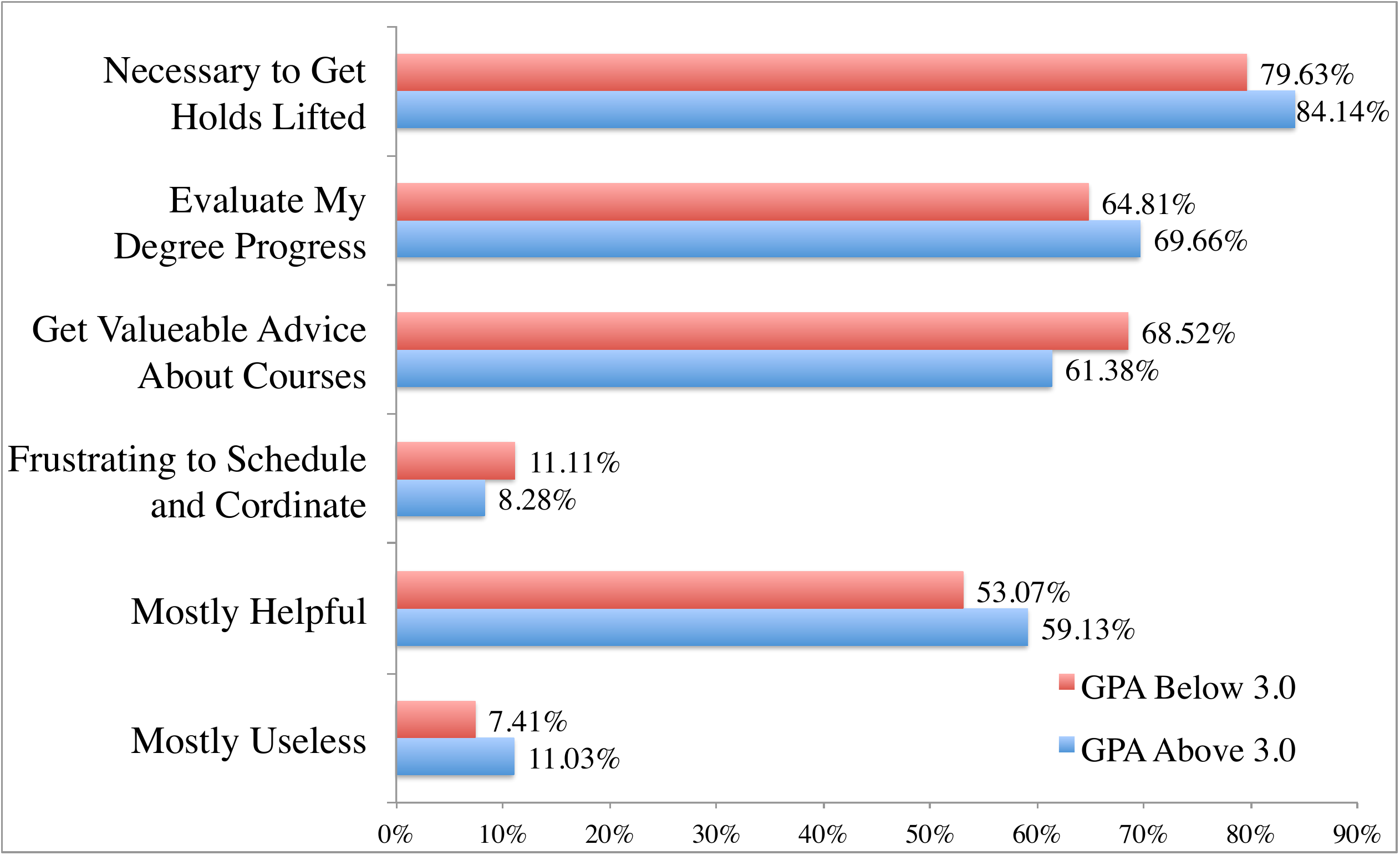}
    \caption{Student answers to the question, ``I view visiting my advisors as...'' broken down
    by those students with a 3.0 or better GPA ($N = 145$) and those below a 3.0 GPA ($N = 54$).
    Students were allowed to select multiple options.}
    \label{fig:visit}
  \end{minipage}
\hfill
  \begin{minipage}[t]{0.48\linewidth}
    \centering
     \includegraphics[height=5cm]{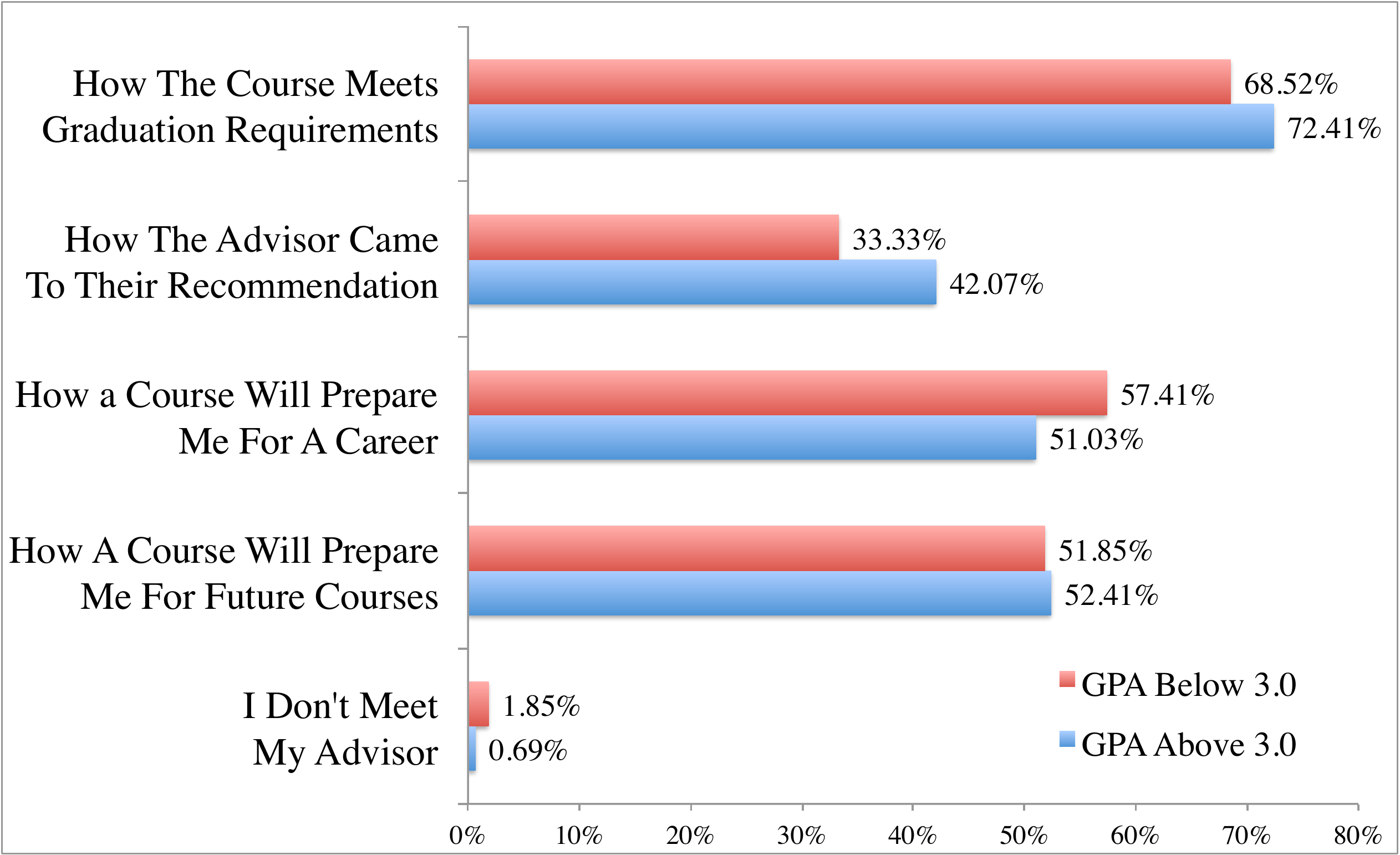}
    \caption{Student answers to the question, ``When receiving advice from an advisor, I like that the advisor explain...''
    broken down by those students with a 3.0 or better GPA ($N = 145$) and those below a 3.0 GPA ($N = 54$).
    Students were allowed to select multiple options.}
    \label{fig:like}
  \end{minipage}
\end{figure*}

\begin{figure*}[ht]
  \begin{minipage}[t]{0.48\linewidth}
    \centering
   \includegraphics[height=5cm]{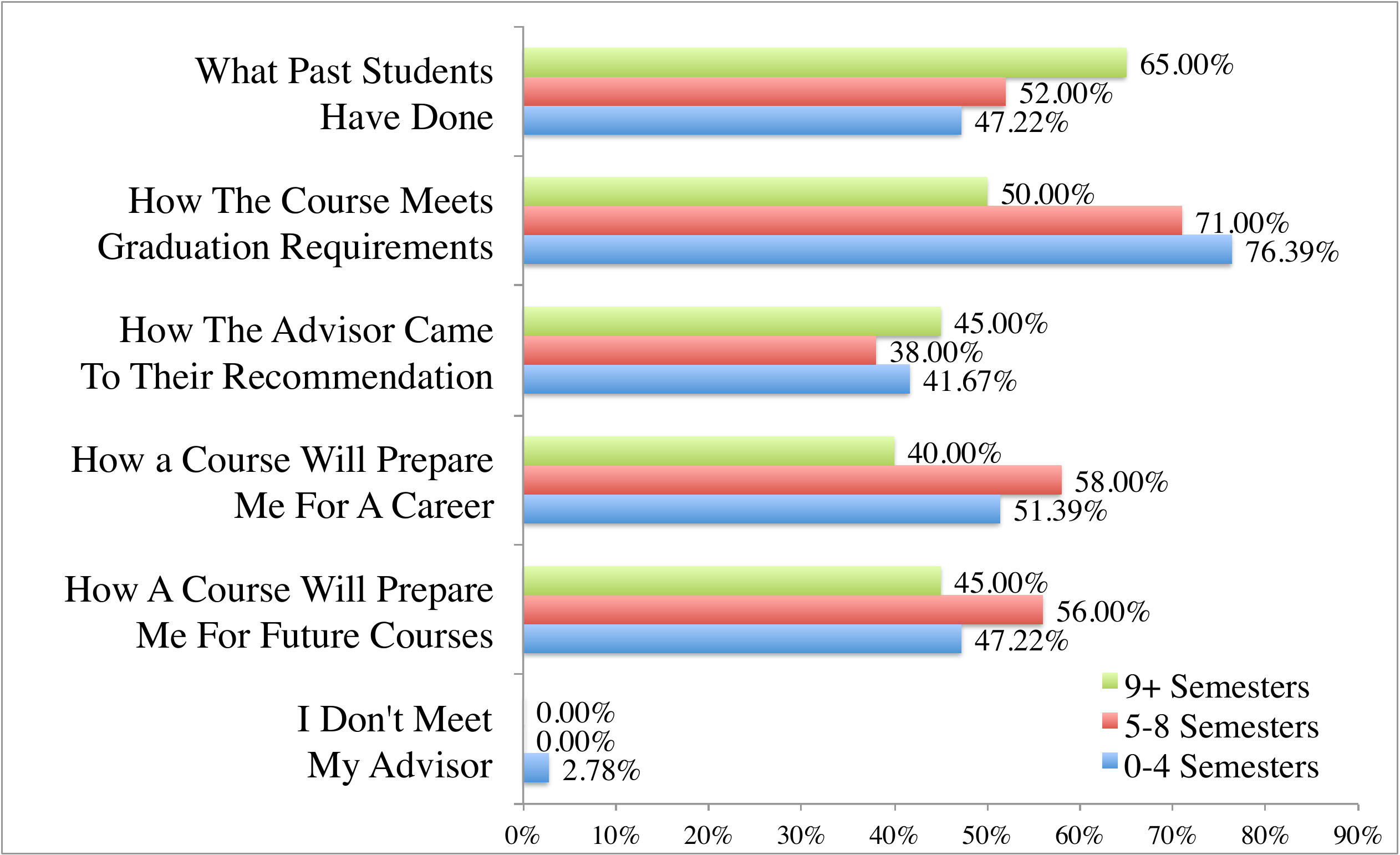}
    \caption{Student answers to the question, ``When receiving advice from an advisor, I like that the advisor explain...''
     broken down by those students attending a tertiary institution for 0-4 semesters ($N = 70$), 5-8 semesters ($N = 100$), and 9+ semesters ($N = 20$).  Students were allowed to select multiple options.}
    \label{fig:like-semester}
  \end{minipage}
\hfill
  \begin{minipage}[t]{0.48\linewidth}
    \centering
     \includegraphics[height=5cm]{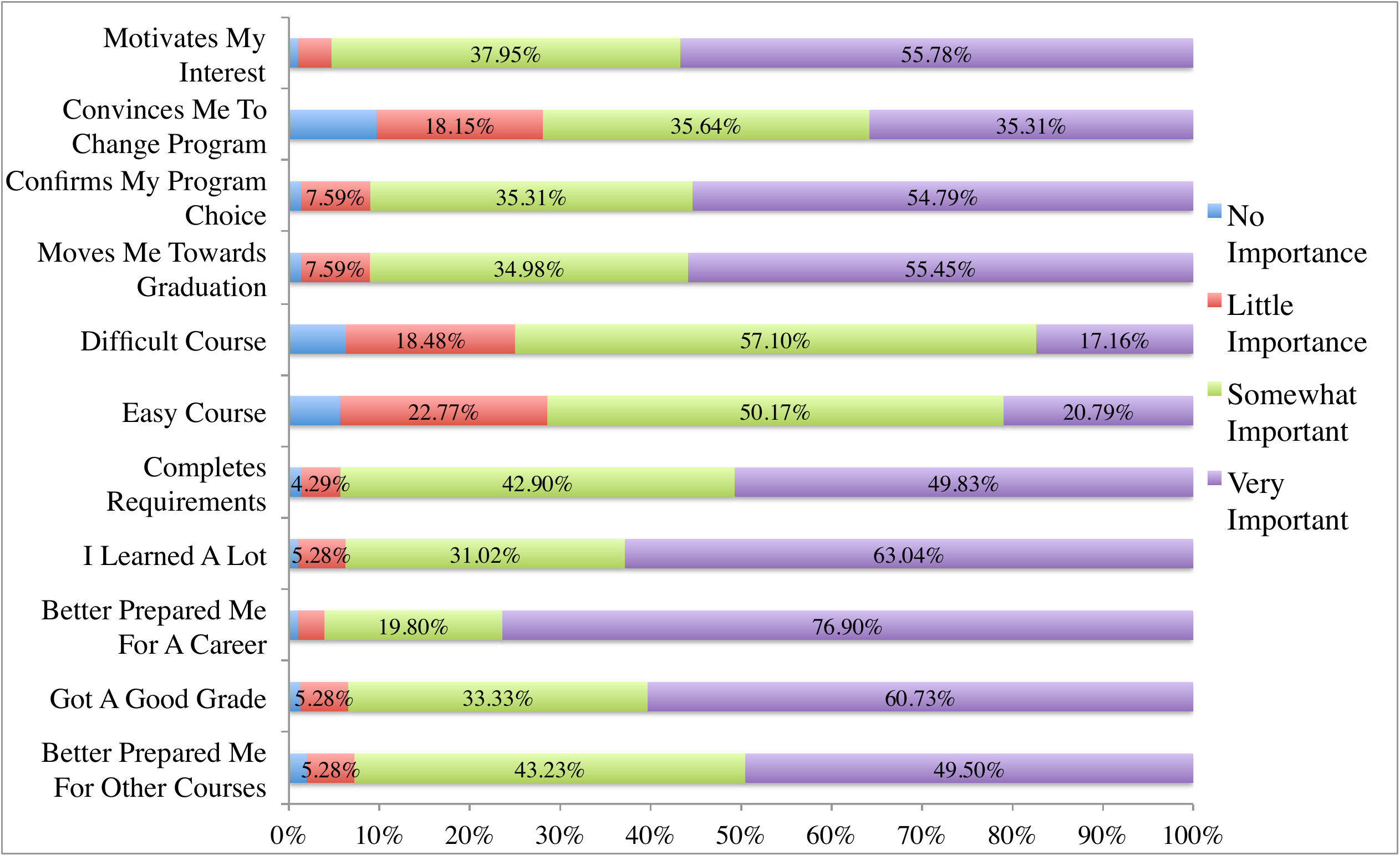}
    \caption{Student answers to the question, ``What factors do you consider important in making a course valuable to you?'' from the AANS survey.
    Students were required to choose one of the options along the scale shown, $N=313$.}
    \label{fig:valuable}
  \end{minipage}
\end{figure*}

\begin{figure*}[ht]
  \begin{minipage}[c]{0.48\linewidth}
    \centering
   \includegraphics[height=5cm]{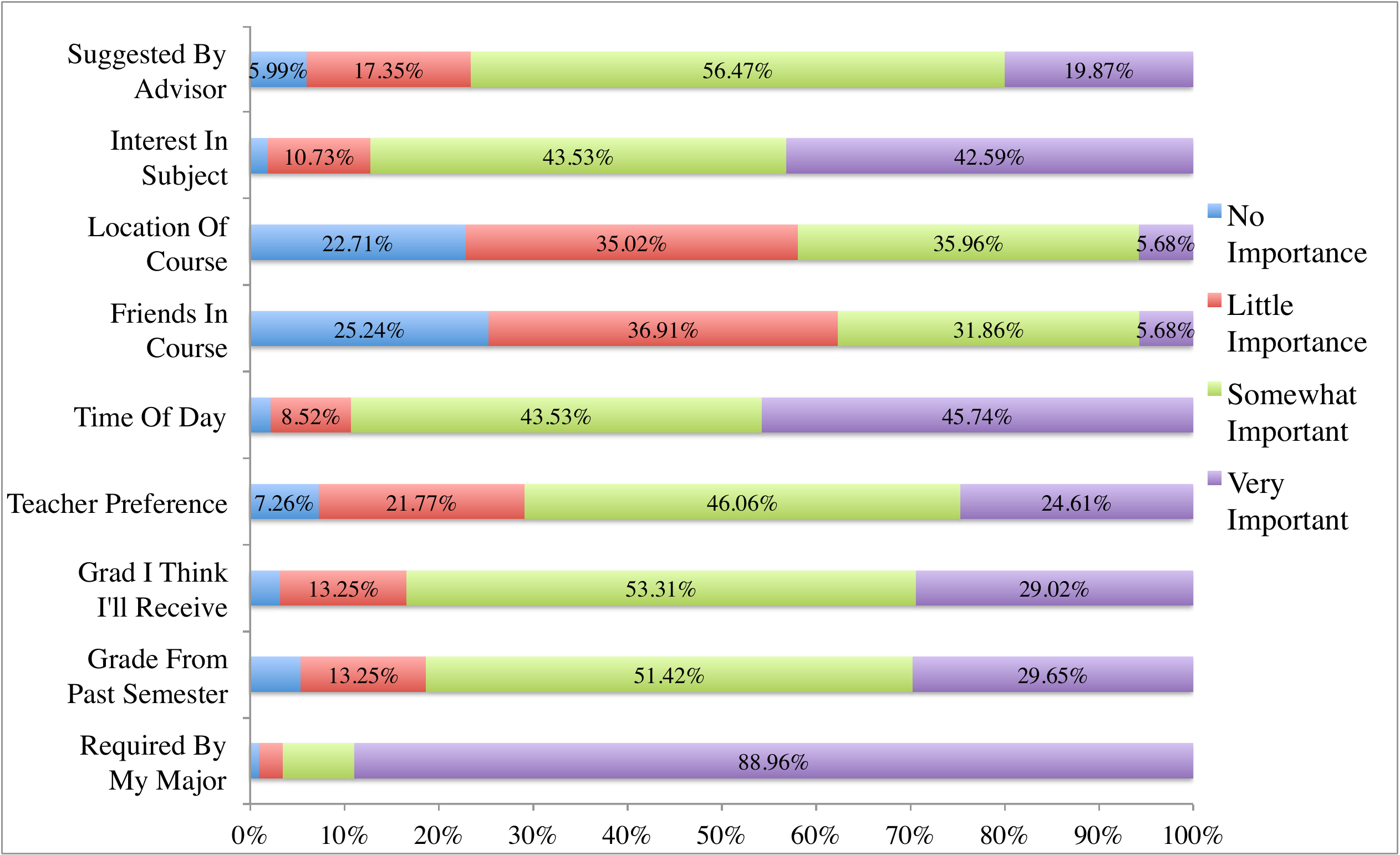}
    \caption{Student answers to the question, ``Please rate the importance
    of the following factors when selecting a course,'' from the AANS survey.  Students were required to choose one of the options
    on along the scale shown, $N=317$.}
    \label{fig:old-pref}
  \end{minipage}
\hfill
  \begin{minipage}[c]{0.48\linewidth}
\newcolumntype{L}[1]{>{\hsize=#1\hsize\raggedright\arraybackslash}X}%
\newcolumntype{R}[1]{>{\hsize=#1\hsize\raggedleft\arraybackslash}X}%
\newcolumntype{C}[1]{>{\hsize=#1\hsize\centering\arraybackslash}X}%
{\small
\begin{tabularx}{\linewidth}{|R{1.5}|C{0.75}|C{0.75}|}
\hline
  							& Borda Score	&	Average List Position	\\ \hline
Required By My Major 						& 855		& 1.70		\\ \hline
Time Of Day 								& 674		& 2.61		\\ \hline
Interest In Subject							& 657		& 2.70		\\ \hline
Professor Teaching The Course				& 580		& 3.09		\\ \hline
Grade I Anticipate Receiving					& 556		& 3.20		\\ \hline
Suggested By My Advisor						& 485		& 3.56		\\ \hline
Grades From Other Courses In Previous Semesters	& 421		& 3.88		\\ \hline
Friends In Course							& 380		& 4.09		\\ \hline
Location Of Course							& 338		& 4.30		\\ \hline
	
\end{tabularx}
}
    \caption{Student responses when asked to rank the elements  in order of importance to them
    when deciding to take a course; unranked elements were assumed to be tied, at the end, 
    of a student's list.  Borda scoring is used to compile the score with the top element receiving 6 points and the last element
    in a list receiving 0 points, $N=200$.}
    \label{fig:rating}
  \end{minipage}
\end{figure*}


\subsubsection{Opinions About Automated Systems}
About $50\%$ of the PSY group and $40\%$ of the CS group wanted 
to work through some ``what if'' scenarios.  These included rearranging 
proposed courses, comparing expected time to graduation for different course selections,
and other factors. If these users had been able to interact with our 
explanation system they could have built and tested these
scenarios in real time, a true benefit of our system.
Additionally, about $10\%$ of students in both groups expressed 
interest in working through whole plans of study for multiple semesters
or entire academic tenures.
Our system currently allows students to walk through their study one semester at a time, sequentially; 
with an appropriate user interface allowing visualization of advice concurrently across multiple semesters,
this is a key area where our system could be of benefit in the future, as the algorithm is explicitly
designed to allow this kind of exploration.

A handful of students (less than $5\%$) asked for more specific learning factors 
that a course would improve and wanted to know how this would translate to their 
future success. In order for our system to answer these questions,
a significantly more complicated model-building process would be
required. This is an example of an area of inquiry where a meeting
with a human advisor would be invaluable --- highlighting where
our system can be used to \emph{encourage} discussion, rather 
than replace it.

There was a small fraction, less than $8\%$ of CS students and no 
PSY students, who wanted to see more numbers and statistics in our system 
instead of our conversational explanations, indicating that
perhaps a minority exists who are more comfortable
reasoning with more objective factors, and whose
concerns are not adequately addressed by the current advising process.

\subsection{Advisor Attitudes}

We surveyed 
10 advisors, including faculty members who perform 
academic advising, advisors attached to a single department, and advisors 
who see students in multiple areas within a single college. 
Our
small sample size does not allow us to present
as a complete a statistical comparison as we would like, but we can still 
draw some conclusions about how advisors view the role of our system.

Nearly all advisors, across all categories, saw requirements as the 
most important priority when recommending courses to students. 
This criteria was rated as the first priority for 9 of 10 advisors surveyed.  
In stark contrast to the students, 7 of 10 advisors rated drawing analogy 
between the current student and past student performance (i.e. case-based reasoning)
as the least 
important aspect of advising.

Advisors rated our data as being generally correct with a median of 
$4.0 / 5.0$ and generally clear with a median of $3.0 / 5.0$.
The advisors saw our advice for the struggling student as less clear and 
less correct because our system did not (and could not) engage the 
student in a discussion about choosing another major.  In fact, when advisors 
did raise issues about the quality of our advice, it was generally in response to 
subjective factors.  Advisors felt that our advice, while technically correct in 
most instances, left out many important factors that could only be addressed
by face-to-face meetings.

The issue of subjective factors was key for the advisors.  They felt that, 
``there is no need to put a computer between two humans that need to 
communicate.''
It was very clear that advisors in our sample were
worried that students, if given access to our system, would skip the person 
to person advising process in favor of a machine---a concern that was
not supported by the results of the student survey.
7 of the 10 advisors said they would rarely or never use our system or 
recommend our system to students.  All three of the advisors who suggested 
giving students access to our system did so with the caveat that students 
should still be required to meet with a human advisor to clear up any 
questions or concerns that the student would have.

The experienced advisors did not always agree with our system, and sometimes 
not with each other.
There was some radically different advice from one advisor to the next given the same proposed
advising situation.  This may be an area where a better understanding of the broad
trends in the student data could support the advice that advisors are generating 
for their students; facilitating advisors to make good decisions, supported by data.

The consensus from the advisors surveyed is that advising is hard.
No two students are the same, and advisors
need to be prepared to direct students to other resources, such as counseling,
testing, and other majors.  The advisors also argued that students are good at figuring out what
courses they want---the advisors' real job is to advise them about subjective
factors such as workload, career preparedness, and setting and achieving realistic goals.

Figures \ref{fig:visit} through \ref{fig:old-pref} show selected results
of our survey in more detail.  In some of these cases we have 
separated out groups of students to compare the attitudes of 
students with higher and lower GPA's and students that are
earlier or later in their academic tenure.

\subsection{Detailed Analysis}
Figure \ref{fig:visit} shows that students with high GPA's were
more negative about the advising products available to them at the time.  
They were more
likely to refer to advising as ``Mostly Useless'' and didn't see advising as
an opportunity to receive information about courses.  However, over 
50\% of both groups were positive about the advising experience.

Figure \ref{fig:like} shows that the high GPA students are more focused on meeting 
requirements and how the advisor came to their recommendation.  One interpretation 
of this data is that higher achieving students are more goal-oriented.  The immediate 
goals of such students could consist of successfully completing courses and earning 
a degree.  Such students appear to be more concerned with satisfying degree requirements 
and understanding how the advice given to them by an advisor may or may not 
directly apply to them, hence the somewhat higher rates of selection for 
\textit{``How The Course Meets Graduation Requirements,''} and 
\textit{``How The Advisor Came To Their Recommendation,''} respectively.

Figure \ref{fig:like-semester}  shows students earlier in their career are more focused
on meeting course requirements.  We conjecture that, by later stages, students
know what they need to do and don't perceive this advice as being as important.
Additionally,
students who have been in tertiary education longer place more
value on advisors 
telling
them what other students in the past have done.
Students in the 5--8 semester cohort are concerned about career advice 
while those in the other two groups, we conjecture, have either already figured out what
they are going to do (9+ semesters) or aren't even thinking about it (0--4 semesters).

Figure \ref{fig:valuable} from the AANS survey shows that students are very career focused.  The highest percentage
of students want courses that directly tie to their careers.  Additionally, students want
good grades, but also want to learn a lot in a course. They are somewhat concerned about the 
difficulty or easiness of the course, but much more interested in whether it completes any of their requirements. 
In that way, they are very goal oriented, and it appears that a safe assumption is that the primary goal of most
students is graduation with a good GPA.

Figure \ref{fig:old-pref} from the AANS survey shows the results when students are asked to compare 
the importance of different elements.
 We see that a course being required
is the overwhelmingly most important thing to students when selecting a course.
Following this, perceived grade, time of day, and subject matter interest
are ranked closely as Somewhat Important or Very Important.  
These results are echoed by responses from the ESS survey shown in Figure
\ref{fig:rating},
which shows that students perceive required-ness and time of day
to be most important.

\section{Discussion and Conclusions}

In our surveys and this paper, we asked what explanations make 
academic advice compelling and convincing.  The primary lesson 
for advisors is that this is not one-size-fits-all.
There is clear variability, even within students in two or three 
large intro classes in Computer Science and Psychology, in what students think about when 
choosing classes, and what they want from advisors and advising support software.  

It is clear that many students will use advising support software if it allows 
them to explore ``what-if'' scenarios, and if it provides clear, understandable 
explanations for its recommendations, particularly in terms of other 
students' experiences.  These preferences carry over, we expect, to human 
advising: students appreciate explanations that begin, ``Many 
students who have had similar grades in these specific courses 
have gotten these grades in this course.''
This result is of particular interest, since it was a technique the advisors did not
like to use.  We conjecture that they feel it de-emphasizes the
uniqueness of the individual student or implicitly sets
expectations which may cause the student to become discouraged when
he or she is not able to ``live up'' to previous students' performances.

Another strong finding is that students want to see the longer-term impacts 
of course choices with respect to their particular goals (often, graduation 
in a ``reasonable'' amount of time with a 
high GPA).  What prerequisites are fulfilled by the recommended 
courses, and what chains of dependent courses are begun?  What 
courses will they be better prepared for, directly and indirectly, by the 
recommended courses?  Finally, how will the recommended 
courses prepare them for post-college opportunities, either directly 
or by preparing them for future useful courses?

We noticed that students with higher GPAs had, on average, 
different expectations and desires.
These diverse needs suggest that advisors should 
be flexible in the reasoning they present to support their advice.  Some 
students want reassurance that they are on track for a career, 
and others seem to simply enjoy school.  
In engineering and computer science , it seems that 
educational careers are often framed in terms of job preparation.
While many students find these topics to be important, we also see
that many students value subjective factors such as the 
topics of the courses (90\%), expected workload, professor, and 
(always!) time of day. 

We have not yet explored advising support systems which account for 
subjective factors. Students expressed a desire for information about so-called 
``hidden factors,'' i.e. \emph{how} a course would prepare them for future classes and for 
post-graduation experiences. 
Learning these hidden factors from data,
including course descriptions, student grades, and course evaluations is an intriguing
area for future research. 
Additionally, a corpus of data could be collected on-line by the 
advising support system itself. Such a system would be designed to collect small 
amounts of preference information from students during the workflow of the 
tool, without being so intrusive as to ask students to complete survey-style questionnaires.

We were surprised to learn that the students 
surveyed focused on more easily evaluated factors 
such as time to graduation and GPA.  These 
two factors will provide an important, albeit
incomplete, basis for more personalized explanations.
However, it is also clear that the weights 
students put on high grades versus time to graduation versus other 
subjective factors depend strongly on the individual.
While some of this information can be gleaned 
from standardized teaching and course evaluations as well as 
word of mouth, the value of human 
advisors is that they can offer their own subjective 
evaluation of course difficulty, popularity, etc. tailored to the goals and abilities of the particular student.

Advising support software, such as the system presented in this paper, 
can help both students and human advisors by allowing the student to perform rote 
requirement checking, as well as providing a platform for informed exploration of course 
choices and possible outcomes, which could facilitate more productive in-person advising sessions
by reducing the amount of time spent on the more formulaic aspects of academic advising.
We reiterate that we are exploring an advising \emph{support} system, and that we found no 
evidence that e-advisors could or should replace human advisors.
Indeed, based on responses from students in Engineering and Computer Science, we found that
they are \emph{more} likely to distrust an e-advisor than non-engineering students.
Responses from students and
advisors underscore the importance of having a
human in the loop for creative
problem solving, subjective analysis, deep understanding of their university or college
system, and most of all, the personal attention that good advisors offer.
We hope that our initial findings about what students want from
the advising process offer something useful to academic advisors.

\section*{Acknowledgements}
Most of the work on this project occurred while all the authors were at 
the University of Kentucky.
We would like to thank Elizabeth Mattei for her help with the statistical analysis,
Jennifer Doerge for her helpful feedback on advising (and generally being an
example of a great advisor), and the members of the UK AI--Lab, especially Robert Crawford,
Daniel Michler, and Matthew Spradling for their support and helpful discussions.  
We are also grateful to the anonymous reviewers
who have made many helpful recommendations for the improvement of this paper.

\bibliographystyle{IEEEtran}
\bibliography{ASEE-explain}

\begin{thebibliography}{10}
\providecommand{\url}[1]{#1}
\csname url@samestyle\endcsname
\providecommand{\newblock}{\relax}
\providecommand{\bibinfo}[2]{#2}
\providecommand{\BIBentrySTDinterwordspacing}{\spaceskip=0pt\relax}
\providecommand{\BIBentryALTinterwordstretchfactor}{4}
\providecommand{\BIBentryALTinterwordspacing}{\spaceskip=\fontdimen2\font plus
\BIBentryALTinterwordstretchfactor\fontdimen3\font minus
  \fontdimen4\font\relax}
\providecommand{\BIBforeignlanguage}[2]{{%
\expandafter\ifx\csname l@#1\endcsname\relax
\typeout{** WARNING: IEEEtran.bst: No hyphenation pattern has been}%
\typeout{** loaded for the language `#1'. Using the pattern for}%
\typeout{** the default language instead.}%
\else
\language=\csname l@#1\endcsname
\fi
#2}}
\providecommand{\BIBdecl}{\relax}
\BIBdecl

\bibitem{DodsonNaturalLanguage2011}
T.~Dodson, N.~Mattei, and J.~Goldsmith, ``A natural language argumentation
  interface for explanation generation in {M}arkov decision processes,'' in
  \emph{Proceedings of the 2nd International Conference on Algorithmic Decision
  Theory (ADT-11)}, 2011, pp. 42--55.

\bibitem{DodsonNatural2013}
T.~Dodson, N.~Mattei, J.~T. Guerin, and J.~Goldsmith, ``An english-language
  argumentation interface for explanation generation with markov decision
  processes in the domain of academic advising,'' \emph{ACM Transactions on
  Interactive Intelligent Systems}, vol.~3, 2013.

\bibitem{gordon2011academic}
V.~N. Gordon, W.~R. Habley, and T.~J. Grites, \emph{Academic advising: A
  comprehensive handbook}.\hskip 1em plus 0.5em minus 0.4em\relax Wiley, 2011.

\bibitem{lowenstein2005if}
M.~Lowenstein, ``If advising is teaching, what do advisors teach?''
  \emph{NACADA Journal}, vol.~25, no.~2, p.~65, 2005.

\bibitem{KingAcademicAdvisingRetention1993}
M.~C. King, ``Academic advising, retention, and transfer,'' \emph{New
  Directions for Community Colleges}, vol. 1993, no.~82, pp. 21--31, 1993.

\bibitem{MetznerPerceivedQuality1989}
B.~S. Metzner, ``Perceived quality of academic advising: The effect on freshman
  attrition,'' \emph{American Educational Research Journal}, vol.~26, no.~3,
  pp. 422--442, 1989.

\bibitem{LoweGivers2001}
A.~Lowe and M.~Toney, ``Academic advising: Views of the givers and takers,''
  \emph{Journal of College Student Retention}, vol.~2, no.~2, pp. 93--108,
  2001.

\bibitem{TitleyNeglectedDimension1982}
R.~W. Titley and B.~S. Titley, ``Academic advising: The neglected dimension in
  designs for undergraduate education,'' \emph{Teaching of Psychology}, vol.~9,
  no.~1, pp. 45--49, February 1982.

\bibitem{econ-gap}
``Minding the gap,'' The Economist, November 16th 2013.

\bibitem{TaltonScavengerHunt2006}
J.~O. Talton, D.~L. Peterson, S.~Kamin, D.~Israel, and J.~Al-muhtadi,
  ``Scavenger hunt: computer science retention through orientation,'' in
  \emph{Proceedings of the 37th SIGCSE Technical Symposium on Computer Science
  Education}, 2006, pp. 443--447.

\bibitem{ISU97Retention}
\BIBentryALTinterwordspacing
C.~Moller-Wong and A.~Eide, ``An engineering student retention study,''
  \emph{Journal of Engineering Education}, vol.~86, no.~1, pp. 7--15, 1997.
  [Online]. Available:
  \url{http://dx.doi.org/10.1002/j.2168-9830.1997.tb00259.x}
\BIBentrySTDinterwordspacing

\bibitem{HartmanLeavingEngineering}
\BIBentryALTinterwordspacing
H.~Hartman and M.~Hartman, ``Leaving engineering: Lessons from {R}owan
  {U}niversity's {C}ollege of {E}ngineering,'' \emph{Journal of Engineering
  Education}, vol.~95, no.~1, pp. 49--61, 2006. [Online]. Available:
  \url{http://dx.doi.org/10.1002/j.2168-9830.2006.tb00877.x}
\BIBentrySTDinterwordspacing

\bibitem{NCSURetention2002}
\BIBentryALTinterwordspacing
L.~E. Bernold, J.~E. Spurlin, and C.~M. Anson, ``Understanding our students: A
  longitudinal-study of success and failure in engineering with implications
  for increased retention,'' \emph{Journal of Engineering Education}, vol.~96,
  no.~3, pp. 263--274, 2007. [Online]. Available:
  \url{http://dx.doi.org/10.1002/j.2168-9830.2007.tb00935.x}
\BIBentrySTDinterwordspacing

\bibitem{godfrey2010retention}
E.~Godfrey, T.~Aubrey, and R.~King, ``Who leaves and who stays? retention and
  attrition in engineering education,'' \emph{Engineering Education}, vol.~5,
  no.~2, pp. 26--40, 2010.

\bibitem{BeaubouefHighAttrition05}
\BIBentryALTinterwordspacing
T.~Beaubouef and J.~Mason, ``Why the high attrition rate for computer science
  students: some thoughts and observations,'' \emph{SIGCSE Bull.}, vol.~37,
  no.~2, pp. 103--106, Jun. 2005. [Online]. Available:
  \url{http://doi.acm.org/10.1145/1083431.1083474}
\BIBentrySTDinterwordspacing

\bibitem{FeghaliWebBased2011}
T.~Feghali, I.~Zbib, and S.~Hallal, ``A web-based decision support tool for
  academic advising,'' \emph{Educational Technology {\&} Society}, vol.~14,
  no.~1, pp. 82--94, 2011.

\bibitem{advising:netflix:10}
J.~R. Young, ``The {N}etflix effect: When software suggests students'
  courses,'' \emph{Chronicle of Higher Education}, April 10 2010.

\bibitem{CF-advising}
S.~Ray and A.~Sharma, ``A collaborative filtering based approach for
  recommending elective courses,'' in \emph{Proc. 5th International Conference
  on Information Intelligence, Systems, Technology and Management (ICISTM
  2011)}, 2011, pp. 330--339.

\bibitem{khan:mse_book}
O.~Khan, P.~Poupart, and J.~Black, ``Automatically generated explanations for
  {M}arkov decision processes,'' in \emph{Decision Theory Models for
  Applications in Artificial Intelligence: Concepts and Solutions}, 2011, pp.
  144--163.

\bibitem{RecSysHandbook_Explanation}
N.~Tintarev and J.~Masthoff, ``Designing and evaluating explanations for
  recommender systems,'' in \emph{Recommender Systems Handbook}, F.~Ricci,
  L.~Rokach, B.~Shapira, and P.~B. Kantor, Eds.\hskip 1em plus 0.5em minus
  0.4em\relax Springer, 2011, pp. 479--510.

\bibitem{mahmood:mdp_travel}
T.~Mahmood, F.~Ricci, and A.~Venturini, ``Improving recommendation
  effectiveness: Adapting a dialogue strategy in online travel planning,''
  \emph{Information Technology \& Tourism}, vol.~11, no.~4, 2009.

\bibitem{tandk:uncertainty}
A.~Tversky and D.~Kahneman, ``Judgement under uncertainty: Heuristics and
  biases,'' \emph{Science}, vol. 185, pp. 1124--1131, 1974.

\bibitem{shani:mdp_bookstore}
\BIBentryALTinterwordspacing
G.~Shani, D.~Heckerman, and R.~I. Brafman, ``An {MDP}-based recommender
  system,'' \emph{Journal of Machine Learning Research}, vol.~6, pp.
  1265--1295, Dec. 2005. [Online]. Available:
  \url{http://dl.acm.org/citation.cfm?id=1046920.1088715}
\BIBentrySTDinterwordspacing

\bibitem{guerin:tr:model}
J.~T. Guerin and J.~Goldsmith, ``Contructing a dynamic {B}ayes net model of
  academic advising,'' in \emph{Proceedings of the Uncertainty in Artificial
  Intelligence Workshop on Bayesian Modeling Applications}, 2011.

\bibitem{guerin2012academic}
J.~T. Guerin, J.~Hanna, L.~Ferland, N.~Mattei, and J.~Goldsmith, ``The academic
  advising planning domain,'' in \emph{Proceedings of the 3rd Workshop on the
  International Planning Competition (WS-IPC-12)}, 2012, pp. 1--5.

\bibitem{nugent:cbe}
C.~Nugent, D.~Doyle, and P.~Cunningham, ``Gaining insight through case-based
  explanation,'' \emph{Journal of Intelligent Information Systems}, vol.~32,
  pp. 267--295, 2009.

\end{thebibliography}
\end{document}